%% file: ijcai18.tex
\pdfoutput=1
\documentclass{article}
\usepackage{ijcai18}

\usepackage{times}
\usepackage{xcolor}
\usepackage{soul}
\usepackage[utf8]{inputenc}
\usepackage[small]{caption}

\usepackage{amsmath,epsfig}
\usepackage{amssymb}
\usepackage{epstopdf}
\usepackage{graphicx,subfigure}
\usepackage{enumerate}
\usepackage{amscd}
\usepackage{rotating}
\usepackage{multirow}
\usepackage{url}
\usepackage{extarrows}
\usepackage{booktabs}
\usepackage{times}
\usepackage{balance}
\usepackage{color}
\usepackage{colortbl}
\usepackage{booktabs} 
\usepackage{array}
\usepackage{qtree}
\usepackage{pgfplots}
\usepackage{mathtools}

\usepackage[linesnumbered,vlined,ruled]{algorithm2e}

\newtheorem{Definition}{Definition}
\newtheorem{Example}{Example}
\newtheorem{Lemma}{Lemma}

\newtheorem{Proof}{Proof}
\newcommand\encircle[1]{%
  \tikz[baseline=(X.base)]
    \node (X) [draw, shape=circle, inner sep=0] {\strut #1};}

\definecolor{orange}{rgb}{1,.5,0}
\definecolor{purple}{rgb}{.85,.44,.84}
\definecolor{color1}{RGB}{255,255,204}
\definecolor{color2}{RGB}{204,255,255}
\definecolor{color3}{RGB}{255,204,204}
\usepackage[utf8]{inputenc}
\title{An Efficient Density-based Clustering Algorithm for Higher-Dimensional Data}

\author{
Thapana Boonchoo$^{\#,\dag}$,
Xiang Ao$^{\#,\dag}$,
Qing He$^{\#,\dag}$,
\\
$^\#$ Institute of Computing Technology, CAS, Beijing 100190, China\\
$^\dag$ University of Chinese Academy of Sciences \\
\{b.thapana, heq\}@ics.ict.ac.cn,
aoxiang@ict.ac.cn
}
\begin{document}

\maketitle

\begin{abstract}
 DBSCAN is a typically used clustering algorithm due to its clustering ability for arbitrarily-shaped clusters and its robustness to outliers.
 Generally, the complexity of DBSCAN is $O(n^2)$ in the worst case, and it practically becomes more severe in higher dimension.  Grid-based DBSCAN is one of the recent improved algorithms aiming at facilitating efficiency. However, the performance of grid-based DBSCAN still suffers from two problems: neighbour explosion and redundancies in merging, which make the algorithms infeasible in high-dimensional space. In this paper, we propose a novel algorithm named GDPAM attempting to extend Grid-based DBSCAN to higher data dimension. In GDPAM, a bitmap indexing is utilized to manage non-empty grids so that the neighbour grid queries can be performed efficiently. Furthermore, we adopt an efficient union-find algorithm to maintain the clustering information in order to reduce redundancies in the merging. The experimental results on both real-world and synthetic datasets demonstrate that the proposed algorithm outperforms the state-of-the-art exact/approximate DBSCAN and suggests a good scalability.

\end{abstract}\vspace{-3mm}


\input{intro}
\input{preliminary}
\input{method1}
\input{method2}
\input{exp}

\input{survey}
\input{conclusion}

\bibliographystyle{named}
\bibliography{acmart}

\end{document}

%% file: intro.tex
\section{Introduction}
Clustering is one of the essential building blocks of data mining.
DBSCAN~\cite{Ester96adensity-based}, as a representative density-based approach, is one of the most widely used algorithms~\cite{Guha:1998:CEC:276304.276312,Pal:2005:PFC:2234548.2235272,Wang:1997:SSI:645923.758369,Xu2015,hartigan1979algorithm,Fisher1987} since its capability of discovering clusters with arbitrary shapes~(over methods such as k-means~\cite{hartigan1979algorithm} which typically returns ball-like clusters).
As a result, DBSCAN and its extensions have been studied for decades and have great contributions to various domains of applications such as spatial database analysis~\cite{Sander:1998:DCS:593419.593465}, forest and land fire indicator analysis~\cite{HERMAWATI2016317}, spam identification~\cite{5529221}, etc~\cite{7929999,1674-4527-14-2-004}.

DBSCAN resorts two parameters to characterize ``density'', namely $\varepsilon$~(a positive real number) and $\mathit{MinPTS}$~(a positive integer). Given a $d$-dimensional data object $p$, the ball centered at $p$ with radius of $\varepsilon$ is considered as \emph{dense} if it covers at least $\mathit{MinPTS}$ objects.
Then the clusters formed by considering all the objects in a dense ball centered at p should be added to the same cluster as p.
Furthermore, two clusters can be merged when the center object of a dense ball is added to another cluster. The merging will be performed to the effect's fullest extent until no more clusters can be merged.

DBSCAN suffers from time-intensive computations since it needs to perform $n$ $\varepsilon$-range queries and cluster labeling propagation for all the objects.
\cite{Gan:2015:DRM:2723372.2737792} pointed out that even efficient data indexing such as r*-tree~\cite{Beckmann:1990:RER:93597.98741} or kd-tree~\cite{Bentley:1975:MBS:361002.361007} is used to index the objects, the complexity of DBSCAN is still $O(n^2)$.
Due to its high time complexity, there are many research efforts devoting to improve the performance of DBSCAN.

Grid-based DBSCAN~\cite{Ade:Thesis:2013,Gan:2015:DRM:2723372.2737792,Sakai2017} is one of the state-of-the-art efficient algorithms that can produce the same clustering result as the original DBSCAN.
The basic idea of Grid-based DBSCAN is to divide the whole dataset into equal-sized square-shaped grids with the side width of $\varepsilon/\sqrt{d}$, where $d$ denotes the data dimension.
Such partition ensures that any two objects in the same grid are within distance $\varepsilon$ from each other. Then the algorithms perform clustering based on neighbour grid query and merging them instead of $\varepsilon$-range queries and cluster labeling propagation.


However, we argue that Grid-based DBSCAN algorithms still suffer from the following two problems. First, the number of neighbour grids increases exponentially with the number of dimensions. Specifically, the number of neighbour grids of a given grid will be $O((2\lceil{\sqrt{d}}\rceil + 1)^{d})$ in the worst case. We provide a formal proof for it in Lemma~\ref{lem:maxnb} in Section~\ref{sec:ourproposed}. Although $O((2\lceil{\sqrt{d}}\rceil + 1)^{d})$  is usually considered as a constant with regard to $n$, it may have a significant impact on the overall performance as the data dimension increases. For example, such number will be more than $10^{20}$ when $d=20$ such that we cannot simply neglect it. We name this problem as \emph{neighbour explosion}.
Second, we observe \emph{symmetry} and \emph{transitivity} in the merging of neighbour grids which enable finer management strategies for such process. However, the existing Grid-based DBSCAN algorithms focus on utilizing efficient solutions to perform the merging rather than optimizing merging management strategies. Recall that the number of neighbour grids increases exponentially with regard to data dimension. Hence we argue that an effective merging management of neighbour grids is in urgent need especially when the data dimension increases because generally we need to check every neighbour grid of a given grid to decide whether they can be merged or not.
Hereafter, we will refer to the Grid-based DBSCAN algorithms as \emph{grid-based algorithms}, unless explicitly stated otherwise.

In this paper, we aim at alleviating the above-mentioned shortcomings of grid-based algorithms and propose an efficient approach, named GDPAM~(\textbf{G}rid-based \textbf{D}BSCAN with \textbf{PA}rtial \textbf{M}erge-Checkings).
The approach utilizes bitmap-like structure to index non-empty grids in multi-dimension so that neighbour grid query can be performed efficiently. For the neighbour grid merging, we devise an effective management strategy that adopts union-find algorithm to maintain the cluster information for pruning unnecessary merging computations. The contributions are summarized as follows.

\begin{itemize}
   \item We propose GDPAM algorithm to extend DBSCAN to higher dimension. Specifically, a bitmap-like indexing called \textbf{H}yper\textbf{G}rid \textbf{B}itmap~(HGB for short) is adopted to index non-empty grids for efficient neighbour grid query.
       Moreover, we propose an effective management strategy that prunes unnecessary merging computations via union-find algorithm on grid topology.
   \item We evaluate the performance of GDPAM on two real-world and four synthetic datasets in different spaces~(ranging from 3D to 54D). The experimental results demonstrate that the proposed GDPAM: 1) outperforms the state-of-the-art grid-based algorithms by three orders of magnitude in high-dimensional spaces, 2) significantly reduces the redundancies in the neighbour grid merging and 3) suggests a good scalability.
\end{itemize}

The rest of this paper is organized as follows. We introduce preliminaries used in this paper in Section~\ref{sec:preliminary}. Section~\ref{sec:ourproposed} presents our proposed algorithm.  The experimental results and discussions are provided in Section~\ref{sec:expr}. We survey the related work in Section~\ref{sec:relatedwork} and conclude the paper in Section~\ref{sec:conclusion}.

%% file: preliminary.tex
\section{Preliminaries}\label{sec:preliminary}
In this section, we first review the original DBSCAN and then introduce the grid-based algorithms.
\subsection{DBSCAN}
\begin{figure} [h!]
\centering
\includegraphics[width=5.5cm, height=2.3cm]{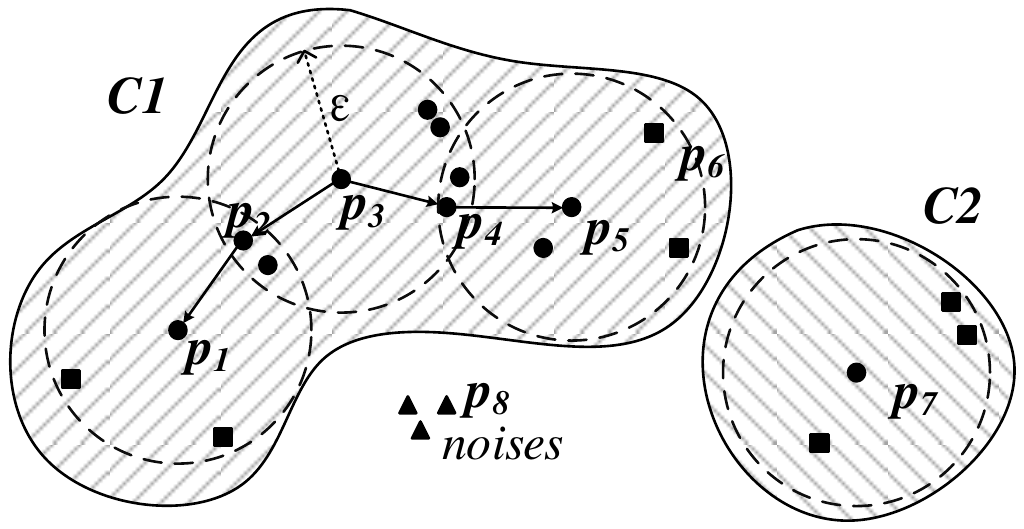}\vspace{-3mm}
\caption{An example of dataset layout. The dashed circles indicate the $\varepsilon$-radii of objects, and let $\mathit{MinPTS}=4$. }\label{example:dbscan}
\end{figure}

DBSCAN groups $d$-dimensional objects with regard to two parameters, $\varepsilon$ and $\mathit{MinPTS}$, where $\varepsilon$ specifies the longest possible distance from an object to its neighbours, and $\mathit{MinPTS}$ specifies the minimum density of subset to form a cluster. Formally, an object in $d$-dimensional space is denoted by $p \in \mathbb{R}^{d}$, and we denote a dataset of objects as $\mathbb{D}$. We use Fig.~\ref{example:dbscan} as a toy example, in which $\varepsilon$ is specified in the figure and $\mathit{MinPTS}$ is set to four. Let $p$ and $q$ be two objects in $\mathbb{D}$.
A neighbour set of $p$, denoted by $\mathrm{N_{\varepsilon}}(p)$, contains objects $q'$ $\in$ $\mathbb{D}$ where $\mathrm{dist}(p,q') \le \varepsilon$. An object $p$ is a \emph{core object} if $|\mathrm{N_{\varepsilon}}(p)|$ $\ge$ $\mathit{MinPTS}$. For example, all the circle objects are core objects in Fig.~\ref{example:dbscan}.
$p$ is \emph{directly density-reachable} from $q$ if $q$ is a core object and $p \in \mathrm{N}_{\varepsilon}(q)$.
For example, $p_6$ is directly density-reachable from $p_5$. Additionally, $p$ is \emph{density-reachable} from $q$ if there exists an object sequence $p_{1}, p_{2}, ..., p_{k}$ where $p_{1} = q $ and $p_{k} = p$ such that $p_{i+1}$ is directly density-reachable from $p_{i}$, where $1 \le i \le k-1$. For instance, $p_1$ is density-reachable from $p_3$ in the running example. $p$ is \emph{density-connected} to $q$ if there exists an object $o$ such that both $p$ and $q$ are density-reachable from $o$, e.g., $p_1$ is density-connected to $p_5$ because of the existence of $p_3$. Finally, a \emph{cluster} $C$ is defined as a non-empty subset of $\mathbb{D}$ such that 1) $\forall p, q$ and $q$ is density-reachable from $p$ w.r.t. $\varepsilon$ and $\mathit{MinPTS}$, if $ p \in C $, then $q \in C$~\emph{(Maximality)}; and 2) $\forall p, q \in C$, $p$ and $q$ are density-connected w.r.t. $\varepsilon$ and $\mathit{MinPTS}$~\emph{(Connectivity)}. A \emph{noise} is an object that is not included in any clusters.
For example, there are two clusters in Fig.~\ref{example:dbscan}, namely $C1$ and $C2$, and $p_8$ is a noise.

\subsection{Grid-based DBSCAN}\label{sec:gridbased}\vspace{-1.0mm}
Gunawan~\cite{Ade:Thesis:2013} proposed the 2D grid-based algorithm with genuine $O(n \log n)$ time. Gan et al.~\cite{Gan:2015:DRM:2723372.2737792} recently extended
the similar idea to higher dimensionality, which consists of four steps: partitioning step, labeling step, merging step and noise/border object identification step, respectively. Partitioning step divides the whole dataset into equal-sized square-shaped grids, each grid has side width of $\varepsilon/\sqrt{d}$. This partitioning ensures the distances of every pair of objects resided in the same grid are at most $\varepsilon$.
Then, the labeling step identifies whether each grid is a core grid by the following definition.


\begin{Definition}\label{def:coregrid}\vspace{-.3mm}
Core grid: A grid with at least one object is called a non-empty grid. A non-empty grid is called a \emph{core grid} if and only if there are at least $\mathit{MinPTS}$ objects inside or there exists at least one core object inside the grid.
\end{Definition}\vspace{-.5mm}

The grid which is labeled as a core grid can form an individual cluster. Two or more core grids can possibly be merged to the same cluster. Hence, the merging step then needs to be carried out. Such step will produce a graph $G$ in which vertices represent core grids, and an edge will be added between two vertices if they can be merged together. Given two core grids $g_1$ and $g_2 $, they can be merged if and only if there exists core objects $p \in g_1$ and $q \in g_2$ such that $\mathrm{dist}(p, q) \le \varepsilon$. Clusters are formed from the fully connected sub-graphs of $G$. 
%
Finally, all non-core objects need to be identified whether they are noises or the border objects of certain clusters in the border/noise object identification step. 

%% file: method1.tex
\section{GDPAM Algorithm}
\label{sec:ourproposed}
In this section, we detail the proposed GDPAM algorithm. We first discuss the shortcomings of grid-based algorithms and then introduce our approach.

\begin{figure} [h!]
\centering
\includegraphics[width=8.5cm, height=4.5cm]{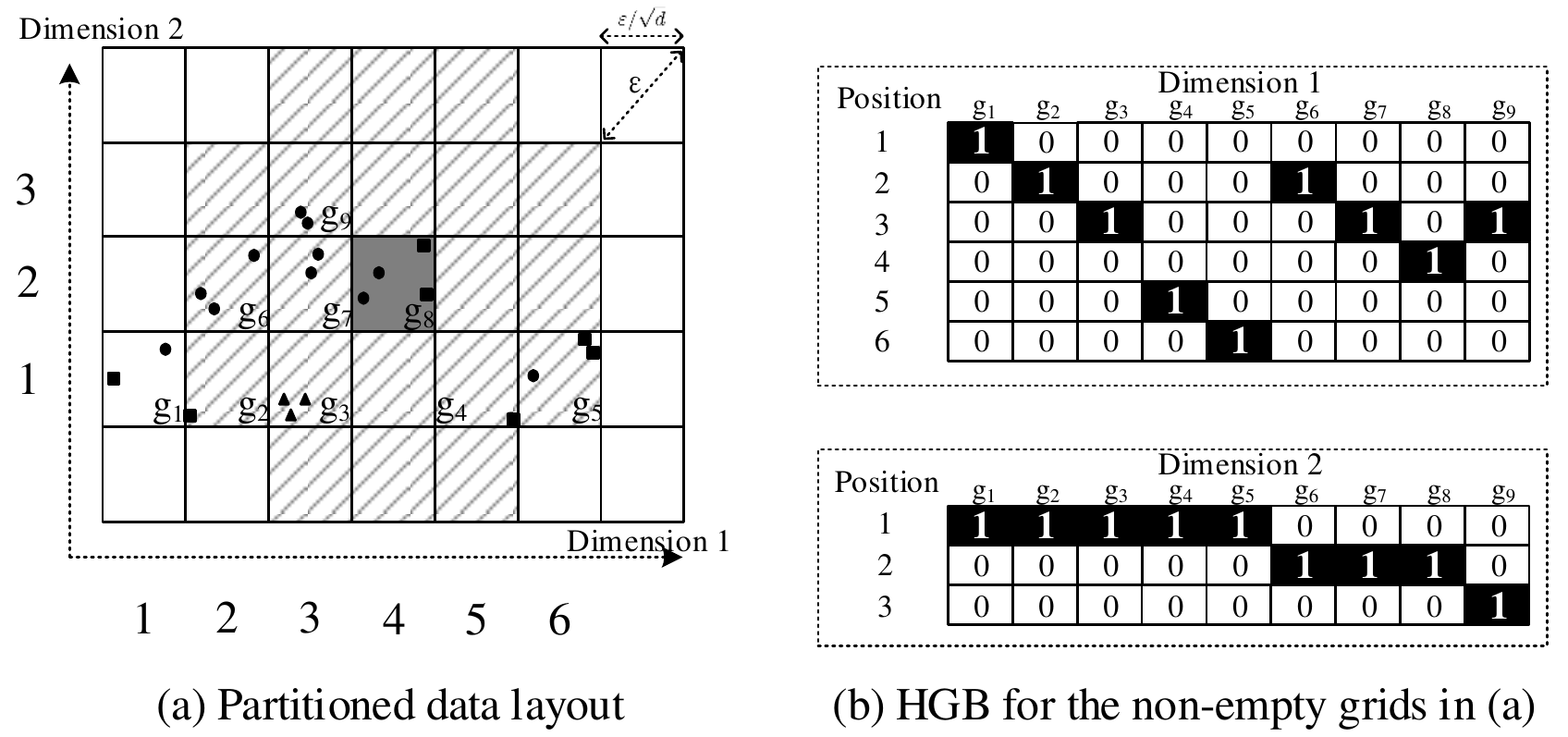}\vspace{-3mm}
\caption{Indexing an example 2D data with HGB structure.}\label{example:data-layout}
\end{figure}
\vspace{-1.0em}
\subsection{Motivation and Framework of GDPAM}
In grid-based algorithms, we find that the number of neighbour grids of a given grid increases exponentially as the dimension increases, and we have the following lemma.
\begin{Lemma}
\label{lem:maxnb}
Consider a $d$-dimensional space, the number of the neighbour grids which needs to be checked for a given grid is $(2\lceil{\sqrt{d}}\rceil + 1)^{d}$ in the worst case.
\end{Lemma}

\begin{Proof}
Without loss of generality, we denote a specific grid in the $d$-dimensional space as $g$. For each direction of individual dimension, the number of neighbour grids of $g$ needs to be checked, denoted by $\mu$ $\in$ $\mathbb{N}$, is calculated as follows:

\begin{displaymath}
\mu \frac{\varepsilon}{\sqrt{d}} \le  \varepsilon, \mu \le \sqrt{d} = \lceil{\sqrt{d}}\rceil.
\end{displaymath}

Since there are two directions for each individual dimension, we need to consider $2\lceil{\sqrt{d}}\rceil+1$ grids for each dimension. Here the number $1$ denotes the grid $g$ itself. Finally, combine all the dimensions together, the number of neighbour grids that the algorithm needs to consider is $(2\lceil{\sqrt{d}}\rceil+1)^{d} - (\tau + 1)$ for every given grid $g$, where $\tau + 1$ is a constant indicating the number of corner grids and $g$ itself which are not included in the set of neighbour grids.
\end{Proof}

According to Lemma \ref{lem:maxnb}, given a specific grid, the running time of its neighbour grid query in the grid-based algorithms will be $O((2\lceil{\sqrt{d}}\rceil+1)^{d})$ in the worst case. For example, the neighbour grids of $g_8$~(the centered grid) in Fig.~\ref{example:data-layout}~(a) are those 20 line-shaded grids.

We name the problem that the number of neighbour grids increases exponentially as \emph{neighbour explosion}. Thus, an efficient indexing technique is demanded to support neighbour grid queries, especially when $d$ increases. As a consequence, we adopt a bitmap-like structure HGB in GDPAM to index non-empty grids such that it supports fast range queries. 

Moreover, we find that the merging step has redundancies that are described as follows. Denoted by $g_1$, $g_2$ and $g_3$ as core grids, we may observe the following redundancies during the merging step.
\begin{itemize}
    \item \textbf{Redundancy 1 (symmetry)}. $g_1$ needs to perform a merging operation with grid $g_2$ and vice versa. Both of them are equivalent, and either one can be skipped.
    \item \textbf{Redundancy 2 (transitivity)}. Assume the $g_1$ and $g_2$ are in the same cluster and so do $g_2$ and $g_3$. We should omit the merging operations between $g_1$ and $g_3$.
\end{itemize}

However, these redundancies are neglected by conventional grid-based algorithms in low data dimension.
It is worth noting that these redundancies can become more severe due to the neighbour explosion problem in higher dimensional space.
The reason is that the more neighbour grids, the more overlap area becomes. Therefore, the grids having overlap neighbours can possibly encounter the above redundancies.
Hence we devise a management strategy in the merging step of GDPAM to alleviate the observed redundant computations.


\textbf{The framework of GDPAM.} The framework of grid-based algorithms carries over to GDPAM almost verbatim. GDPAM also consists of the four steps, and the only differences are the way we index the non-empty grids~(detailed in Section~\ref{sec:bitmapdef}) and the way that we perform the merging step~(detailed in Section~\ref{sec:merge}). 

\subsection{Indexing with HGB}
\label{sec:bitmapdef}
HGB (HyperGrid Bitmap) is a bitmap-like structure used to index non-empty grids in the GDPAM framework. It is composed by $d$ two-dimensional bit arrays, where $d$ denotes the data dimension. We denote each of the bit array as $\mathcal{B}_{i}$ where $1\leq i \leq d$.
Each $\mathcal{B}_{i}$ can be regarded as a table with $k_{i}$ rows and $N_{g}$ columns, where $k_i$ is the number of distinct positions of grids that contain objects in $i$-th dimension, and $N_{g}$ denotes the number of non-empty grids.
In other words, each $\mathcal{B}_i$ is used to record the position of every non-empty grid in $i$-th dimension.
To set up each $\mathcal{B}_i$, we initially set all bits of it to $0$, and encode each grid from id $1$ to $N_{g}$.
Considering a non-empty grid whose id = $x$, where $1 \leq x \le N_{g}$, we first find the position of that grid in $i$-th dimension, denoted as $j$ for example, and set the corresponding bit of $\mathcal{B}_{i}$, i.e., $\mathcal{B}_{i}[j,x]$ to $1$. $\mathcal{B}_{i}[j,x]$ $=$ $1$ indicates the position of the grid $x$ in $i$-th dimension is $j$.

\begin{Example}
Fig.~\ref{example:data-layout}~(b) shows the HGB for the toy example in Fig.~\ref{example:data-layout}~(a). In the example, we have 9 non-empty grids in $2$D space, thus $N_{g}$ $=$ $9$ and $d$ $=$ $2$. As a result, we need $2$ bit arrays, namely $\mathcal{B}_{1}$ and $\mathcal{B}_2$, such that each of which contains 9 columns.
Meanwhile, $\mathcal{B}_{1}$ and $\mathcal{B}_{2} $ contain $6$ and $3$ rows since they have $6$ and $3$ distinct valid positions, respectively.
For a specific grid, e.g. $g_8$ (id=$8$), we have $\mathcal{B}_{1}[4, 8]=1$ and $\mathcal{B}_{2}[2, 8]=1$ as shown in Fig.~\ref{example:data-layout}~(b).
\end{Example}
Once HGB is completely constructed, we can simply perform a conventional range query on HGB of a given grid $g$ by setting the range for each dimension as follows:

\begin{displaymath}
\mathrm{NeighbourRange}(g, i) = [g.pos[i] - \lceil{\sqrt{d}}\rceil, g.pos[i] + \lceil{\sqrt{d}}\rceil],
\end{displaymath}
where $i$ denotes $i$-th dimension. 
We show how we perform a neighbour grid query on HGB by the following example. Note that for simplicity here we do not exclude the corner grids in the neighbour query of the example.
\begin{Example}
In Fig.~\ref{example:data-layout} consider a neighbour grid query of $g_8$, the ranges of dimension 1 and dimension 2 are [2, 6] and [1, 3], respectively. Then slices collected from tables corresponding to dimension 1 and dimension 2 are $\mathcal{B}_{1}$[2:6,:] and $\mathcal{B}_{2}$[1:3,:], respectively. First we perform bitwise operations \verb"OR" on $\mathcal{B}_{1}$[2:6,:] and $\mathcal{B}_{2}$[1:3,:], the outputs are [0, 1, 1, 1, 1, 1, 1, 1, 1] and [1, 1, 1, 1, 1, 1, 1, 1, 1], respectively. Then the final bitwise operation \verb"AND" is performed on the two bit arrays, outputting [0, 1, 1, 1, 1, 1, 1, 1, 1]. Thus, the set of neighbour grids of $g_8$ is all the grids except $g_1$.
\end{Example}

%

\subsubsection{Complexity Analysis}
\label{sec:space}
First, we analyze the space complexity of HGB. Recall that we have $d$ bit arrays in HGB, where $d$ is the data dimension. For simplicity, we assume every bit array has $\kappa$ distinct positions.  Since each bit array has $\kappa$ rows and $N_g$ columns~(the number of non-empty grids), the space complexity is thus $O(d\cdot\kappa\cdot N_{g})$. Second, the time complexity of constructing HGB is $O(d \cdot N_{g})$, since we need to scan all the non-empty grids and set the corresponding bits of bit arrays for every dimension. For neighbour query operation, the time complexity is $O(d \cdot (2\lceil{\sqrt{d}}\rceil + 1))$ $=$ $O(d^{3/2})$, because it needs to collect slices~(of the range size $(2\lceil{\sqrt{d}}\rceil + 1))$ of the $d$ bit arrays.

\subsection{Merging Management Strategy}
\label{sec:merge}

Our merging management strategy makes use of union-find algorithm to alleviate the redundancies.  In the \emph{union-find} algorithm, we can perform two basic operations, i.e., $\mathrm{Union(\cdot)}$ and $\mathrm{Find(\cdot)}$. Basically, $\mathrm{Find(e)}$ returns the representative of subset that the element $e$ belongs to. And, $\mathrm{Union(e_{1}, e_{2})}$ merges the two subsets that $e_{1}$ and $e_{2}$ belong to. 

\subsubsection{Grid-based DBSCAN with Partial Merge-Checkings}
\begin{figure} [!htpb]
\centering\vspace{-3mm}
\includegraphics[width=8.75cm, height=2.5cm]{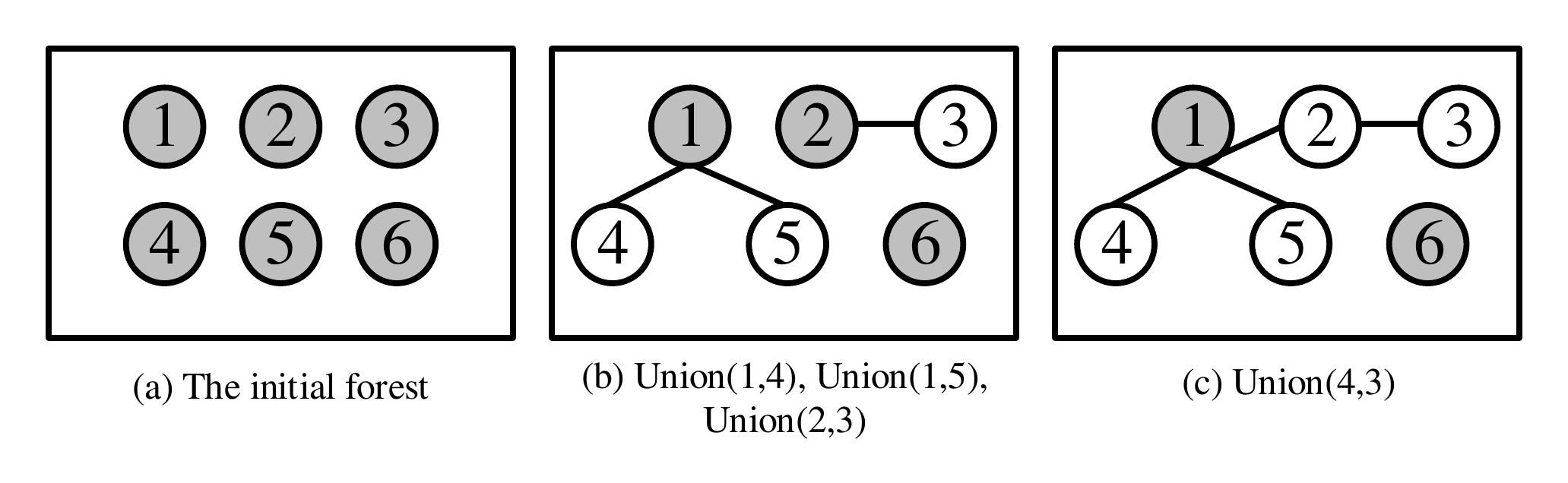}\vspace{-5mm}
\caption{Running example of the forest during the merging step}\label{example:forest}
\end{figure}

In GDPAM, we use a forest-like structure to maintain the clustering information in the merging step. Initially, every core grid is considered as an individual tree (primary cluster), and our strategy is to merge those trees if they meet the merge-checking criteria (refer to Section~\ref{sec:gridbased}). Finally, after all core grids have been processed, we obtain the final forest such that core grids residing in the same tree result in being in the same cluster.

\begin{algorithm}\scriptsize
\caption{Merging Strategy}\label{alg:GDPAM}
\KwIn{ $\mathcal{B}$: HGB for the dataset, $\mathcal{C}$: Set of core grids}

\KwOut{$\mathcal{F}$: Forest produced by GDPAM}
    Initial the trees in $\mathcal{F}$ as one-node trees, each tree corresponds to a core grid.\label{GDPAM:l1}\\
    \ForEach{core grid $g$ $\in$ $\mathcal{C}$}{\label{fun2:l3}
        $N$ $\gets$ neighbour grids query of $g$ on HGB\\
        \ForEach{$g'$ $\in$ $N$}{\label{fun2:l3}
            \If{$g'$ is a core grid}{
                \If{$\mathcal{F}.\mathrm{Find(g)}$ == $\mathcal{F}.\mathrm{Find(g')}$}{\label{GDPAM:l6}
                    Skip merge-checking between $g$ and $g'$\label{GDPAM:l7}\\
                }
                \Else{
                    \If{$g$, $g'$ can be merged (Section \ref{sec:gridbased})}{\label{GDPAM:l9}
                        $\mathcal{F}.\mathrm{Union(g,g')}$\label{GDPAM:l10}\\
                    }
                }
            }

        }
    }
    \Return $\mathcal{F}$
\end{algorithm}

The pseudo-code of the merging strategy is depicted in Algorithm~\ref{alg:GDPAM}. First, the forest initially consists of $N_c$ trees, where $N_c$ is the number of core grids~(Line \ref{GDPAM:l1}). For example, Fig.~\ref{example:forest}~(a) shows the initial forest of the $6$ core grids. Note that in the example a circle node corresponds to a core grid, and a shaded node denotes a root. Then, each grid will perform merge-checkings with their neighbour core grids. Prior to that, we first test whether the two core grids have already been in the same cluster or not. We invoke $\mathrm{Find(\cdot)}$ in order to return the roots of trees that those two core grids reside in~(Line \ref{GDPAM:l6}). If the returned roots are the same, we can safely skip the merge-checking as the two core grids have already been in the same cluster~(Line \ref{GDPAM:l7}). For example, according to the forest in Fig.~\ref{example:forest}~(b), the core grids \encircle{4} and \encircle{5} \emph{do not} need to perform a merge-checking since they have the same root, i.e., $\mathrm{Find(\encircle{4})}$ $=$ $\mathrm{Find(\encircle{5})}$ = \encircle{1}. Otherwise, we need to perform a merge-checking~(refer to Section \ref{sec:gridbased}). If we find that two core grid can be merged to the same cluster, we perform $\mathrm{Union(\cdot)}$ operation between the two core grids in the forest~(Line \ref{GDPAM:l9}--\ref{GDPAM:l10}). Specifically, the $\mathrm{Union(\cdot)}$ invokes $\mathrm{Find(\cdot)}$ to return the roots of trees that those two core grids reside in, and assigns one of the roots as a child of the other. Fig.~\ref{example:forest}~(c) shows an example of performing $\mathrm{Union(\encircle{4}, \encircle{3})}$. It assigns $\mathrm{Find(\encircle{3})}=\encircle{2}$ as $\encircle{1}$'s child because of $\mathrm{Find(\encircle{4})}=\encircle{1}$.
Recall that the merge-checking is a costly operation, skipping redundant merge-checkings can considerably improve the overall performance. Though the time complexity of GDPAM is the same as conventional grid-based algorithms in the worst case, we can observe a clear time saving of GDPAM in the experiments.

%% file: exp.tex
\section{Experiments}\vspace{-1mm}
\label{sec:expr}
In this section, we present the results of our experimental studies on real-world and synthetic datasets.
\subsection{Experimental Settings}
All the experiments were conducted in a workstation equipped with four Intel (R) CPU E5-2609 v3 processors and 128 GB RAM running a Linux Cent OS 6.5. We implemented our proposed GDPAM with C++.
\subsubsection{Datasets}
We evaluated our algorithm on four synthetic and two real-world datasets. Table~\ref{table:datainfo} depicts their statistics.

\textbf{Synthetic Datasets} We generated the synthetic datasets by a generator URG.
The generator takes 4 parameters: the number of objects ($n$), the number of clusters ($c$), the number of dimensions ($d$), and the percent of noise ($pnoise$) [default=0.0005\%]. We used URG to generate five different kinds of datasets, i.e., 3-, 10-, 15-, 20-, 30-, and 40-dimensional datasets in range 1000 to 10000 in each dimension. To avoid too-dense cluster, when $0.00025n$ objects have been generated, the data will have possibility to move a bit ($33\%$ for $-5, 33\%$ for $+5$) in each dimension. For simplicity and convenience, we denote each of them by its dimensionality when discussing in the following parts. For example, if we set $n=3, c=10, d=3$, the URG will generate a dataset with $3$ million objects grouped into $10$ clusters in $3$-dimensional space, and we denote it as 3D.

\textbf{Real-world Datasets} All the real datasets are obtained from UCI Machine Learning Repository \cite{Lichman:2013}. We evaluated on 7- and 54-dimensional datasets.  For the 7- and 54-dimensional datasets, we follow~\cite{Gan:2015:DRM:2723372.2737792} to use Individual household electric power consumption (Household) and PAMAP2 \cite{Reiss:2012:INB:2357489.2358027} as the $7$-, $54$-dimensional datasets, respectively.


\begin{table}\scriptsize
\centering
\caption{Dataset statistics}\vspace{-3mm}
\label{table:datainfo}
\begin{tabular}{lcllc}
\hline\hline
\textbf{Dataset} & \multicolumn{1}{l}{\textbf{Dimension}} & \textbf{Type} & \textbf{\#Objects} & \multicolumn{1}{l}{\textbf{\#Clusters}} \\ \hline\hline
3D               & 3                                      & Synthetic     & 3,000,000          & 10                                      \\ \hline
10D              & 10                                     & Synthetic     & 3,000,000          & 10                                      \\ \hline
30D              & 30                                     & Synthetic     & 3,000,000          & 10                                      \\ \hline
40D              & 40                                     & Synthetic     & 3,000,000          & 10                                      \\ \hline
Household        & 7                                      & Real          & 2,075,259          & N/A                                     \\ \hline
PAMAP2           & 54                                     & Real          & 3,850,505          & N/A                                     \\ \hline\hline
\end{tabular}
\end{table}

\subsubsection{Compared Methods and Parameter Settings}
The compared methods in the experiments include\vspace{-1mm}
    \begin{enumerate}
       \item DBSCAN: original DBSCAN~\cite{Ester96adensity-based} with r*tree,\vspace{-1mm}
       \item GRID~\cite{Gan:2015:DRM:2723372.2737792}: A state-of-the-art grid-based exact DBSCAN algorithm,\vspace{-1mm}
       \item GRID-A~\cite{Gan:2015:DRM:2723372.2737792}: A state-of-the-art grid-based approximate DBSCAN algorithm,\vspace{-1mm}
       \item HGB: our proposed method with only HGB indexing,\vspace{-1mm}
       \item GDPAM: the full version of our proposed method.\vspace{-1mm}
    \end{enumerate}
For the implementation of DBSCAN, GRID, and GRID-A, we used the binary code which is implemented by C++ and publicly available\footnote{ \url{https://sites.google.com/site/junhogan/}.}. 
We investigated the datasets and followed the suggestions produced by the parameter selection tool~\cite{Sander1998} for setting $\varepsilon$ and $\mathit{MinPTS}$ with regard to the range and dimension of datasets, as well as the number of grids.
\subsection{Experimental Results}
In this subsection, we demonstrate the experimental results. Note that all the reported execution time of the compared methods is a 3-time-run average value. In addition, we did not include some experimental results of DBSCAN because it failed to report the results within 15 hours.

\subsubsection{\textbf{Overall Performance}}

\begin{figure*}[!htpb]
  \centering
  \includegraphics[width=17.0cm,height=2.5cm]{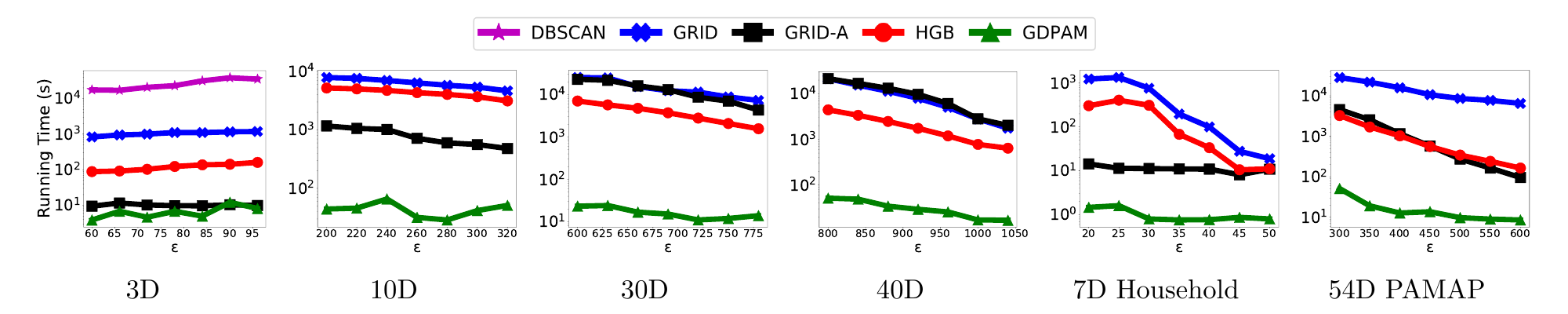}\vspace{-4mm}
  \caption{Running time on synthetic and real-world datasets.}\label{fig:overallperf}
\end{figure*}

Fig.~\ref{fig:overallperf} illustrates the execution time of each compared method on both synthetic and real-world datasets in \textbf{log scale}. 
From the results on the synthetic data shown in Fig.~\ref{fig:overallperf}, first we observed that HGB always runs faster than DBSCAN and GRID. 
For example, HGB runs $197.59\times$ faster than DBSCAN (3D dataset, $\varepsilon=60$, $\mathit{MinPTS}=20$), and almost $6\times$ faster than GRID (40D dataset, $\varepsilon=800$, $\mathit{MinPTS}=80$). Since HGB produces efficient neighbour grid query, the proposed method obtains a clear time-saving. 

Second, GDPAM, with the merging management strategy for reducing redundant computations, significantly outperforms the other compared methods. It achieves approximately $3000\times$ speedup compared with DBSCAN. Moreover, it yet has almost up to three orders of magnitude of speedup compared with GRID and our HGB.
In addition, we surprisingly observed that the GDPAM method is even faster than the recent GRID-A, which is claimed as a $O(n)$ approximate DBSCAN algorithm, when the data dimension is larger than $3$. For example, GDPAM achieves $736\times$ speedup compared with GRID-A (30D dataset, $\varepsilon=600$, $\mathit{MinPTS} = 70$). We argue that the reason for such observation is that the number of the neighbour grids that needs to be checked grows exponentially with the data dimension. As a consequence, we cannot neglect it as the data dimension increases, and our proposed merging management strategy suggests its effectiveness.
Furthermore, Fig.~\ref{fig:overallperf}~(b), Fig.~\ref{fig:overallperf}~(c), and Fig.~\ref{fig:overallperf}~(d), respectively, also show that when the data dimension increases, GDPAM still achieves advantages compared with GRID-A. 

Similar observations can also be found from the results of real-world data shown in Fig.~\ref{fig:overallperf}~(d), Fig.~\ref{fig:overallperf}~(e). That is, HGB is faster than GRID, and GDPAM, moreover, clearly leads the compared two exact algorithms. For instance, HGB runs $15.38\times$ faster than GRID, on the other hand, GDPAM runs $128.52\times$ and $1191.97\times$ faster than HGB and GRID (PAMAP2 dataset (54D), $\varepsilon=400$, $\mathit{MinPTS} = 150$), respectively.

\subsubsection{\textbf{Effectiveness of HGB}}\label{sec:effHGB}
\begin{figure}[!htpb]
  \centering
  \includegraphics[width=8.5cm,height=2.5cm]{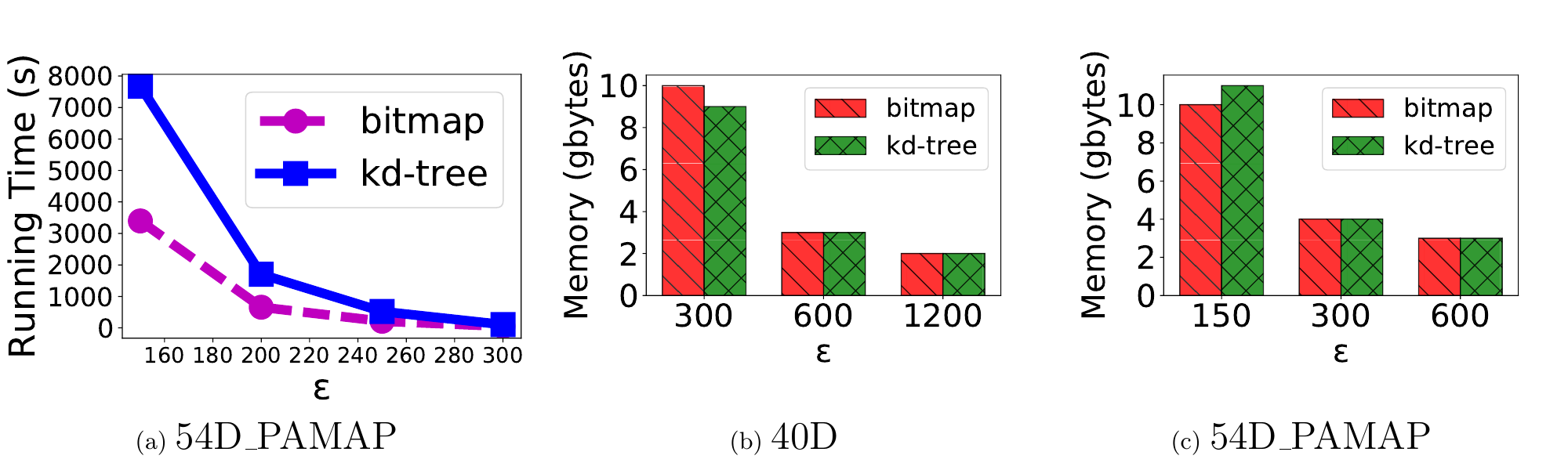}\vspace{-3mm}
  \caption{
  Effectiveness of HGB.}\label{fig:memory} 
\end{figure}

To demonstrate the effectiveness of HGB, we show the running times of our framework with traditional kd-tree and HGB as indexing techniques by fixing $\mathit{MinPTS}$ and varying $\varepsilon$ in Fig.~\ref{fig:memory}~(a). We can observe that querying neighbour grids using HGB runs $2\times$ faster than that of kd-tree when $\varepsilon$ is small~($\varepsilon$ $=$ $150$, the number of the neighbour grids increases). Note that small $\varepsilon$ challenges the clustering process as the number of partitioned grids will significantly increase.
In addition, Fig.~\ref{fig:memory}~(b), Fig.~\ref{fig:memory}~(c) show the memory consumptions of kd-tree and HGB indexing on 40D synthetic and 54D real-world datasets, respectively. We can observe that kd-tree and bitmap have almost a similar amount of consumed memory in most cases. We conclude that the bitmap indexing is more efficient than kd-tree for indexing non-empty grids in terms of execution time, while their  memory consumptions are not significantly different. 


\subsubsection{\textbf{Effectiveness of Merging Management Strategy}}

\begin{figure}[!htpb]
  \centering
  \includegraphics[width=3.5cm,height=1.8cm]{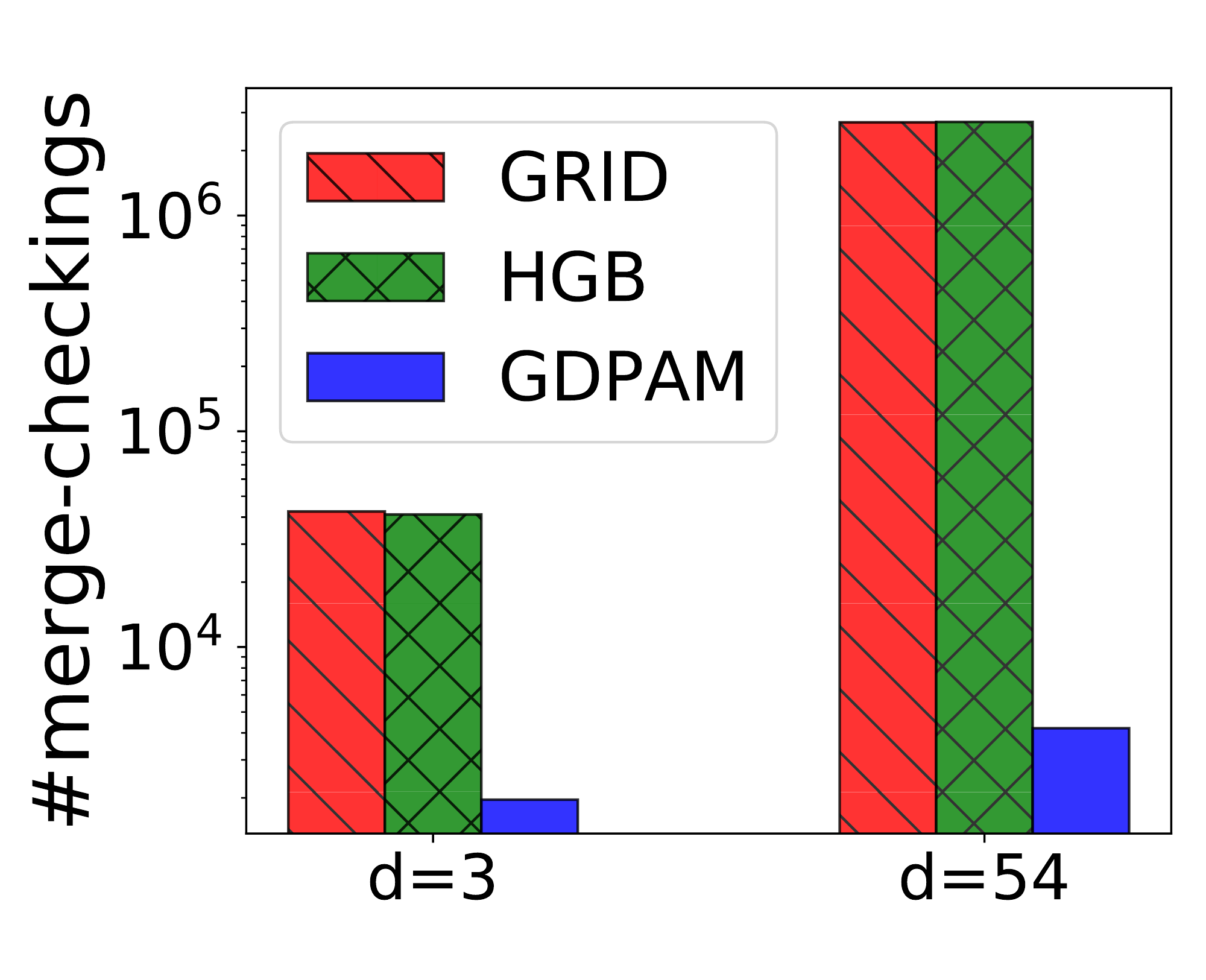}\vspace{-3mm}
  \caption{
  Effectiveness of GDPAM.}\label{fig:mergecounts} 
\end{figure}

Next we visualized the number of merging operations as shown in Fig.~\ref{fig:mergecounts} to exhibit the effectiveness of the merging management strategy. It is clear that the numbers of operations used by GRID and HGB are almost the same. The reason is that we only use HGB to index non-empty grids without any specialized techniques to avoid merging redundancy.  In addition, GDPAM achieves a significant operation-saving compared with HGB and GRID. For example, GDPAM performs only $0.15\%$, $4.62\%$ of merging operations compared with GRID on 54D real-world and 3D synthetic datasets, respectively. As expected, we can see the saving ratio on 54D dataset is much greater than that of 3D. The reason is that we encounter more redundancies in higher dimension. The results demonstrate the effectiveness of GDPAM in redundancy reducing in the merging step.

\subsubsection{\textbf{Scalability}}
\begin{figure}[!htpb]
  \centering
  \includegraphics[width=8.0cm,height=3.85cm]{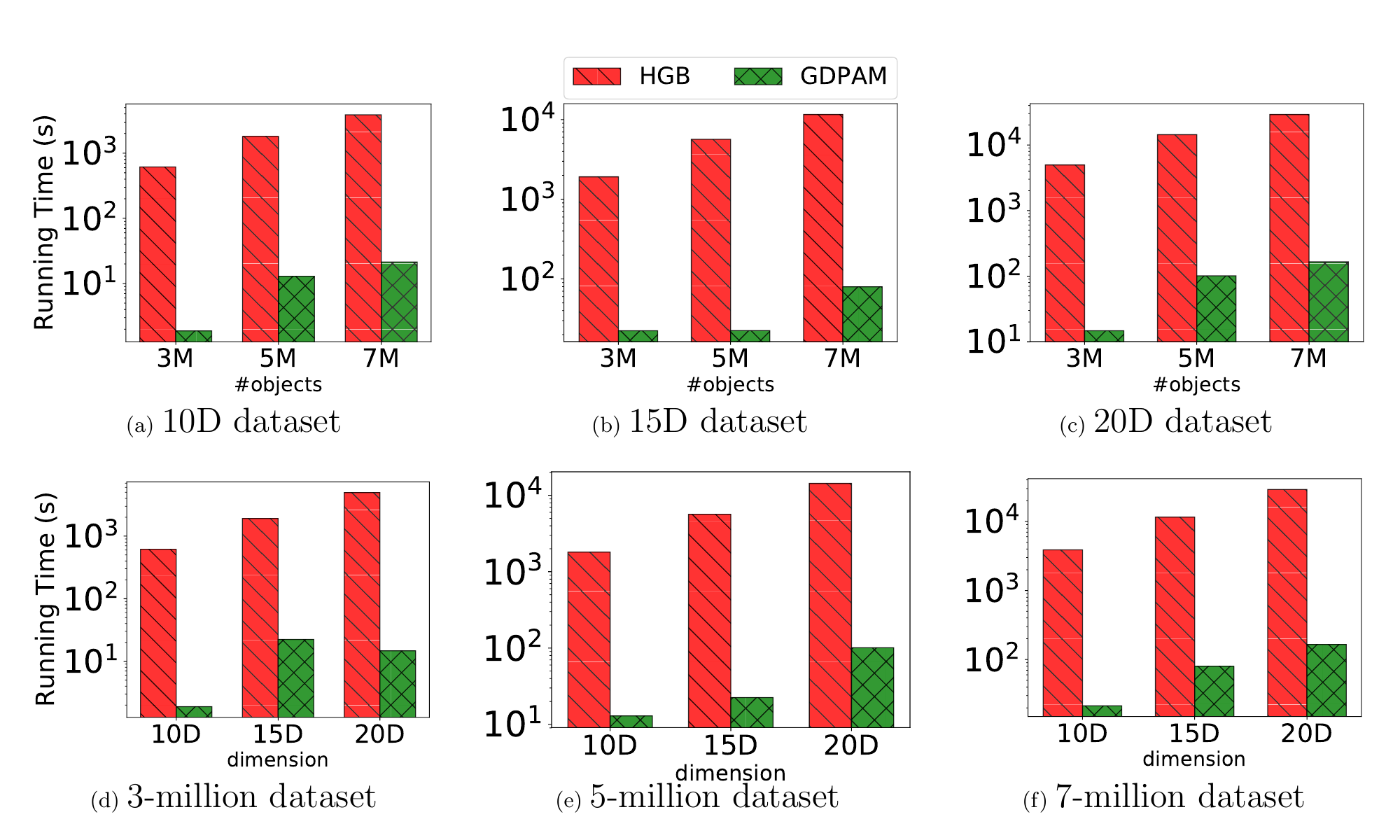}\vspace{-3mm}
  \caption{
  Scalability.}\label{fig:scale} 
\end{figure}

Finally, we examined the scalability of the proposed algorithms as we increased the input size/data dimension.
In particular, we generated datasets by using URG and set $n$ to 3, 5, 7 on each $d$ = 10, 15, 20. Then we obtained nine datasets, and we run both HGB and GDPAM on each of them. First, we visualized the execution time by fixing the data dimension and varying the data size as shown in Fig.~\ref{fig:scale}. Interestingly, from the figure, we view the performance of both HGB and GDPAM that they rise lower than linear increase. Furthermore, GDPAM rises much slower than HGB.
We analyze that though GDPAM theoretically has the same time complexity of merging step as HGB in the worst case, it can still dramatically reduce some of the symmetric and transitive redundant merging operations. Thus, it scales well to large datasets.
We second examined the scalability to the data dimension and showed in the same figure. From the figure, we can find similar observations as the previous investigation that both HGB and GDPAM scale well as the data dimension increases. Additionally, GDPAM is more stable to the variation of data dimension.

%% file: survey.tex
\section{Related work}
\label{sec:relatedwork}
We broadly categorize the related papers in the literature on improving the efficiency of DBSCAN as follows.

\textbf{Sampling-based DBSCAN algorithms} are designed to improve the time-consuming object neighbour query operation in DBSCAN. For example, Zhou et al~\cite{Aoying:2000:ASD:362237.362239} proposed a sampling-based method which selects a few representatives, rather than all the neighbour objects, as seeds to expand the clusters. However, this method possibly misses some objects. Moreover, it still raises a representative selection problem which directly impacts on both performance and accuracy of DBSCAN. Another sampling-based method can be found in~\cite{Tsai:2009:GNE:1576659.1576697}. However, it cannot produce the exact results as well.\vspace{-0mm}

\textbf{Grid-based DBSCAN algorithms.} CIT~\cite{4594646} is the first one that develops grid-based algorithm for DBSCAN. The algorithm partitions the data layout into grids with width greater or equal to $2\varepsilon$ and sets a boundary of neighbour search to $\varepsilon$ around the grids. 
Our work is related to~\cite{Ade:Thesis:2013} and \cite{Gan:2015:DRM:2723372.2737792}. Gunawan proposed~\cite{Ade:Thesis:2013} a 2D grid-based algorithm which can terminate in genuine $O(n\log n)$.
Gan et al.~\cite{Gan:2015:DRM:2723372.2737792} extended the grid-based algorithm to solve DBSCAN in more dimensions. Even though the observation of \cite{Gan:2015:DRM:2723372.2737792} makes a good contribution about the complexity of DBSCAN, recently, there is an interesting disputation \cite{Schubert:2017:DRR:3129336.3068335} on some inaccurate statements of the paper \cite{Gan:2015:DRM:2723372.2737792}. In addition, the authors suggested when and why we should still use original DBSCAN. Sakai et al.~\cite{Sakai2017} proposed an improved grid-based algorithm which refines the merging step by using minimum bounding rectangle criteria. While in this paper we devise a novel grid-based DBSCAN solution with an efficient and compact index as well as an improve merging management strategy, which extends the algorithm to higher dimensional data.

\textbf{Other efficient DBSCAN algorithms.} There are other efforts for improving the efficiency of DBSCAN. Some of them focus on producing distributed or parallel version of DBSCAN~\cite{6121313,6877517,Patwary:2014:PPA:2683593.2683655,7929990}. Approximate algorithm is another kind of method to accelerate DBSCAN~\cite{Gan:2015:DRM:2723372.2737792}. Mai et al.~\cite{Mai:2016:AEA:2939672.2939750} proposed an anytime DBSCAN which employs active learning for learning the current cluster structure, it thus can select only necessary objects to expand clusters.
\vspace{-0.4em}

%% file: conclusion.tex
\vspace{-1.5mm}\section{Conclusions}\vspace{-1.5mm}
\label{sec:conclusion}

In this paper, we have pointed out the state-of-the-art grid-based DBSCAN algorithms suffered from neighbour explosion and merging redundancies, which make them infeasible to higher dimensional data.
To address such problem, we proposed a novel GDPAM algorithm in this paper. In GDPAM, we devise HGB structure to index non-empty grids for efficiently neighbour grid query. Furthermore, GDPAM intergraded a merging management strategy such that we can safely skip unnecessary merging computations. The results on six real-world and synthetic datasets showed the superiority of the proposed method over the state-of-the-art DBSCAN. \vspace{-1.5mm}
